\documentclass[%
 reprint,
nofootinbib,
 amsmath,amssymb,
 aps,
]{revtex4-1}

\usepackage{graphicx}
\usepackage{dcolumn}
\usepackage{bm}
\usepackage[hidelinks]{hyperref}
\usepackage[mathlines]{lineno}
\usepackage[]{cleveref}
\usepackage[english]{babel}
\usepackage{amssymb}
\usepackage{amsmath}
\usepackage{latexsym}
\usepackage{cancel}
\usepackage{xcolor}
\usepackage{esint}

\usepackage[normalem]{ulem}

\def\be#1\ee{\begin{align}#1\end{align}}

\newcommand{\appropto}{\mathrel{\vcenter{
  \offinterlineskip\halign{\hfil$##$\cr
    \propto\cr\noalign{\kern2pt}\sim\cr\noalign{\kern-2pt}}}}}

\begin{document}

\title{Quasi-normal modes and Microscopic Structure of the Schwarzschild Black Hole}

\author{Mariano Cadoni$^{a,b}$}
\email{mariano.cadoni@ca.infn.it}

\author{Mauro Oi$^{a,b}$}
\email{mauro.oi@ca.infn.it}


\author{Andrea P. Sanna$^{a,b}$}
\email{asanna@dsf.unica.it}

\affiliation{$^a$Dipartimento di Fisica, Universit\`a di Cagliari, Cittadella Universitaria, 09042, Monserrato, Italy}
\affiliation{$^b$Istituto Nazionale di Fisica Nucleare (INFN), Sezione di Cagliari, Cittadella Universitaria, 09042 Monserrato, Italy}

\date{\today}

\begin{abstract}
Maggiore  observed that, in the high-damping regime, the quasi-normal modes  spectrum for  the Schwarzschild black hole resembles that of a quantum harmonic oscillator. Motivated by this observation,  we describe a black hole as a statistical ensemble of $N$ quantum harmonic oscillators.
By working in the canonical ensemble, we show that,  in the large-mass black hole limit, the leading contribution to the Gibbs entropy is the Bekenstein-Hawking  term, while the subleading  one is a logarithmic correction, in agreement with several results in the literature.
We also find that  the number of oscillators scales holographically with the area of the event horizon.
\end{abstract}

\maketitle

\textit{Background and motivations}. Classically, a perturbed black hole reacts dynamically, producing characteristic oscillations, called quasi-normal modes (QNMs), which decay exponentially in time. At linear perturbation level, QNMs correspond to complex eigenfunctions of the system, namely modes characterized by complex frequencies, whose imaginary part describes the damping of the mode in time (see, e.g. Refs. \cite{Berti:2009kk, Kokkotas:1999bd, Konoplya:2011qq}). Moreover, boundary conditions at infinity and at the horizon imply a discrete spectrum for the  frequencies  $\omega_n$, with the imaginary part depending on an integer $n$, the overtone number.

In the high-damping regime (large-$n$ limit), the spectrum of QNMs for a Schwarzschild black hole (SBH), with mass $M$, is independent of $l$, the angular momentum ``quantum'' number, and reads \footnote{We adopt natural units, $c=\hbar =1$.} \cite{Schutz:1985km, Nollert:1993zz, Hod:1998vk, Motl:2002hd} (see Refs. \cite{Kokkotas:1999bd, Nollert:1999ji, Berti:2009kk, Konoplya:2011qq} for reviews)
    \begin{equation}
       8\pi G M  \omega_{\text{n}} \sim \ln 3+2\pi \text{i} \left(n+\frac{1}{2} \right) + \mathcal{O}\left(n^{-1/2}\right).
\label{highdampingQNM}
    \end{equation}

By studying the system's response to external perturbations, in principle we  could also have access to its internal microscopic structure. Therefore, despite being classical, QNMs could contain signatures of quantum gravity effects, encoding information about the quantum properties of black holes and their horizons \cite{Agullo:2020hxe}. This is particularly  true in the large-$n$ limit, which is expected to probe  the  black hole at  short distances.
 
There are several indications supporting this  perspective. On one hand, QNMs could  be useful to understand the AdS/CFT conjecture. In fact, in the case of AdS black holes, the damping of QNMs can be mapped into the thermalization of the conformal field theory on the boundary \cite{Horowitz:1999jd}. On the other hand,  the emergent  and corpuscular gravity scenarios  suggest that black holes could be characterized by long-range quantum gravity effects of $N$ quanta building the black hole \cite{Dvali:2011aa, Dvali:2013eja, Casadio:2015lis, Casadio:2016zpl, Cadoni:2020mgb, Casadio:2021cbv}. Similarly to what happens  for the  surface gravity  of a black hole  (see  Ref. \cite{Cadoni:2020mgb}),  the  quantum  nature of Eq. (\ref{highdampingQNM})  is obscured (the Planck constant $\hbar$ does not appear)  by expressing  $\omega_n$  in terms of the black-hole mass $M$, but becomes fully evident when  we express it in terms of the black hole  temperature $T_\text{H}$.

Thus, the QNMs spectrum  could  represent a coarse-grained description of the response of these $N$ microscopic degrees of freedom to external perturbations, in the same spirit as  the spectrum of the black body radiation  is a manifestation of the collective behavior of a photon gas. 

However, the no-hair theorem \cite{Israel:1967za, Hawking:1971vc} makes a black hole drastically different from a black body. In the latter case  the extensive thermodynamic parameters scale with the volume, while the number of photons, i.e. the number of microscopic degrees of freedom, is independent of the size of the system. For a black hole, on the other hand, the mass $M$ and the entropy $S$ are fixed by the temperature or, equivalently, by the horizon radius $r_h$. From a corpuscular gravity point of view, moreover, we can consider the black hole as a macroscopic quantum state  that  saturates a maximally packaging condition \cite{Dvali:2010bf, Dvali:2010jz, Dvali:2011th, Casadio:2021cbv},  which has been shown to be equivalent to the holographic scaling of $N$ \cite{Cadoni:2020mgb}. Essentially, in a black hole of  a given mass, we can ``pack" a maximum amount of degrees of freedom, which is constrained by the size of the system.

The proposal  of using QNMs to capture some microscopic  properties of black holes is not completely new, but was first proposed by Maggiore in \cite{Maggiore:2007nq}: the linear scaling of the QNMs frequencies with  the overtone number $n$ suggests that a SBH can be described, in the high-damping limit, as a harmonic oscillator, with proper frequency 
\begin{equation}
\omega = \sqrt{\omega_{\text{R}}^2 + \omega_{\text{I}}^2},
\label{Maggioreomega}
\end{equation}
 where $\omega_{\text{R}}$ and $\omega_{\text{I}}$ are the real and imaginary parts of the frequencies \eqref{highdampingQNM} respectively \footnote{A similar proposal for the description of a black hole as a harmonic oscillator, from a corpuscular gravity perspective, can be found in \cite{Casadio:2013hja}.}.
 
Until now, the high-damped QNMs spectrum has been used in the quantum gravity  context  to explain and fix the area spectrum of the event horizon (see, e.g. Refs. \cite{Hod:1998vk, Dreyer:2002vy, Kunstatter:2002pj, Corda:2015rna}),  whose quantization was first suggested  by  Bekenstein in \cite{Bekenstein:1974jk, Bekenstein:1995ju}.  This also  allowed to fix  the Barbero-Immirzi parameter \cite{Dreyer:2002vy}, which is essential to correctly account for the Bekenstein-Hawking (BH) entropy in  Loop Quantum Gravity \cite{Immirzi:1996dr, Agullo:2008yv}.

In this letter, we use Maggiore's result to model the SBH as a canonical ensemble of $N$ harmonic oscillators and derive the black-hole entropy using the QNMs frequencies only, without assuming  horizon-area quantization. Consistently with the no-hair theorem, for an asymptotic observer, the only physical observable is the black-hole mass $M$, which also determines the QNMs frequency spectrum. On the other hand, we assume that, quantum mechanically, the horizon area  and the temperature  can fluctuate  independently from $M$. This will allow us to consistently define the canonical ensemble  and to circumvent the no-hair theorem at quantum level.       

\textit{The model}. In the high-damping regime, $\omega_\text{I} \gg \omega_{\text{R}}$, QNMs probe the internal structure of the black hole, as the wavelength of each oscillator gets smaller and smaller  as $n$ grows.
From \cref{highdampingQNM,Maggioreomega}, we easily get the  frequency spectrum
\begin{equation}
\omega_n \simeq |\omega_{\text{I}}| = \frac{1}{4GM} \left(n+\frac{1}{2} \right) + \mathcal{O}(n^{-1/2}).
\label{freqenergy}
\end{equation}

 Following Maggiore's proposal, we model the black hole as a statistical ensemble of $N\gg 1$ indistinguishable non-interacting (at least in a first approximation) quantum harmonic oscillators with  frequencies
\begin{equation}
\omega_n = \omega_0\left(n+\frac{1}{2}\right),
\end{equation}
where $\omega_0=1/4GM $ is the proper frequency of each oscillator. 

Our derivation relies entirely on equilibrium statistical mechanics, without resorting to  usual black-hole thermodynamics. The black hole will be regarded as an ensemble in  thermal equilibrium with its surroundings at temperature $T=1/\beta$. We will therefore work entirely in the canonical ensemble and consider the number of oscillators $N$ fixed. This is  motivated by the no-hair theorem, which tells us that the chemical potential of a SBH is zero, being the mass $M$ the only classical hair of the hole. 
Considering the SBH as a system at fixed temperature is also consistent with the fact that the QNMs spectrum is computed at fixed black-hole mass. For the asymptotic observer, the latter is related to the Hawking temperature, which is therefore fixed. 

It is very important to stress that, in our statistical description, we treat $\beta$ and $\omega_0$  as \textit{independent}  variables,  i.e. the temperature of the ensemble can change independently  from the black-hole mass.
At first sight, this may seem at odds with standard black-hole thermodynamics. However, we argue that quantum mechanically this is fully consistent.

In standard black-hole thermodynamics, the Hawking temperature $T_\text{H}$ can be defined as  the coefficient of proportionality between the entropy (the area of the black-hole event horizon, $\mathcal{A}_\text{H}$) and its energy (the mass $M$). Classically, $\mathcal{A}_\text{H}$ is a function of $M$, i.e. $\mathcal{A}_{\text{H}}=16 \pi G^2 M^2$. The latter, however, should be considered as the mean value, measured at infinity, of the area of the event horizon, which can fluctuate around its expectation value, from a quantum mechanical point of view \cite{Bekenstein:1974jk, Bekenstein:1995ju, Gour:1999ta} \footnote{This is also supported  by  the  fuzzball proposal for black holes in string theory  \cite{Mathur:2005zp}.}. Only local measurements would allow to probe these fluctuations \cite{Gour:1999ta}. An observer at infinity therefore would not have access to them, as the only degree of freedom he/she  can measure is the classical hair of the black hole, i.e. its mass. This tells us that, at least quantum mechanically, the area of the event horizon can fluctuate \textit{independently} from $M$. Hence,   only  the observer at infinity  can make the identification  $\beta=\beta_\text{H}= 1/T_{\text{H}}=8\pi G M$.  

Using the spectrum \eqref{freqenergy}, the statistical Boltzmann weight of each harmonic oscillator black hole microstate is therefore
\begin{equation}
e^{-\beta \omega_n} = e^{-\beta \omega_0\left(n+\frac{1}{2} \right)} e^{-\beta\frac{\kappa}{\sqrt{n}}},
\end{equation}
where $\kappa$ is a dimensionful constant, proportional to $\omega_0$ on dimensional grounds, parametrizing the low-$n$ behavior of the damping modes. As long as we consider the limit of large-mass black holes $\omega_0 \to 0$ (or the high-temperature limit, $\beta \to 0$), the factor $e^{-\beta \kappa n^{-1/2}}$ can be set equal to $1$  and the partition function will be insensible to the subleading terms  $\mathcal{O}(n^{-1/2})$. 

Being the SBH effectively featureless, except from its mass, and having zero chemical potential,   the probability of occupying a given  energy level  will be the same for all oscillators. The partition function for the composite system of $N$ oscillators therefore reads
\begin{equation}
\begin{split}
&\mathcal{Z} = \left(\sum_{n=0}^{\infty} e^{-\beta \omega_0 \left(n+\frac{1}{2} \right)} \right)^{N} = \left(\frac{e^{\beta \omega_0/2}}{e^{\beta\omega_0}-1} \right)^{N},\\
&\ln \mathcal{Z} = \frac{N\omega_0}{2}\beta - N \ln \left(e^{\beta \omega_0}-1 \right).
\end{split}
\end{equation}

Using standard statistical mechanics relations, we compute the mean energy and the entropy
\begin{equation}
\langle E \rangle = -\partial_\beta \ln \mathcal{Z} = \frac{N \omega_0}{2}\text{cotgh}\left(\frac{\beta \omega_0}{2} \right);
\label{meanenergy}
\end{equation}
\begin{equation}
S =\ln \mathcal{Z} + \beta \langle E \rangle = -N \ln \left(e^{\beta \omega_0} -1 \right) + \frac{ N\beta \omega_0 e^{\beta \omega_0}}{e^{\beta \omega_0}-1 }.
\label{statisticalentropy}
\end{equation}
As expected  for consistency, the expressions above satisfy the first law of thermodynamics, $d\langle E \rangle = T dS$.

Let us now focus on  macroscopic  black holes,  by considering the large $M $ limit, i.e.  $\omega_0 \to 0$. By expanding the mean energy \eqref{meanenergy} and the entropy \eqref{statisticalentropy}, we get:
\begin{equation}
\langle E \rangle = \frac{N}{\beta}+\frac{N}{12}\beta \omega_0^2 + \mathcal{O}(\omega_0^3);
\label{energyexpanded}
\end{equation}
\begin{equation}
S = N-N \ln \left(\beta \omega_0\right) + \frac{N}{24} \beta^2 \omega_0^2 + \mathcal{O}(\omega_0^3).
\label{entropyexpanded}
\end{equation}
We see that the leading terms in the  expansions satisfy $\langle E \rangle = T S$, hence they  capture only the purely thermal extensive contribution $TS$ to the mean energy. 
However, for $\beta \to \infty$ (zero-temperature limit),  the sub-leading terms in \cref{energyexpanded} diverge at fixed $\omega_0$. The same problem appears in standard black-hole thermodynamics, where a zero Hawking temperature implies a divergence of the black-hole mass  $M$. However, this is an artefact of the expansion. This divergence problem can be simply solved in our approach by  first taking the $\beta \to \infty$ limit in the exact expression for $\langle E \rangle$ given by \cref{meanenergy}. We get  $\langle E \rangle = N \omega_0/2$. This is a finite value, which cures the divergence appearing in \cref{energyexpanded} and has the simple physical interpretation of the zero-point energies $\omega_0/2$ of the $N$ oscillators,  representing therefore the contribution of the vacuum. Notice that this contribution  cancels out when we perform first the $\omega_0 \to 0$ limit  and keep the leading terms only. 

The black-hole mass $M$ measured by an observer at infinity can be  seen as the sum of the purely extensive contribution of \cref{energyexpanded} and the contribution of the vacuum $E_V$.  Our  microscopic description  of the SBH in terms of $N$ non-interacting harmonic oscillators holds in the large-$n$  limit. Thus,   we  may expect   deviation of $E_V$ from the naive  value $N\omega_0/2$. Owing to the absence of an external scale  different from  the thermal one $\beta_\text{H}=1/T_\text{H}$, we nevertheless   expect $E_V$ to get  only order-one corrections, and $M$  to take the form  
\begin{equation}\label{vacuumcontribmass}
M = \frac{N}{\beta} + c \frac{N \omega_0}{2},
\end{equation}
where $c$ is a  ${\cal O }(1)$ constant, which can be fixed  using symmetry arguments. 
The  expansions  \eqref{energyexpanded} and \eqref{entropyexpanded}  can be also obtained by considering the limit $\beta \to 0$ instead of $\omega_0 \to 0$ in \cref{meanenergy,statisticalentropy}. Despite being mathematically equivalent, these two limits have a very different physical meaning. While the latter corresponds to black holes with large masses, the former is related to small-mass SBHs. Treating $\omega_0$ and $\beta$ separately introduces some kind of duality between small and large black holes, which is again a consequence of the absence of  an external scale  different from $\beta_\text{H}$.  This implies that in our microscopic description there is no difference between the thermodynamic properties of small- and large-mass black holes (a behavior very different from AdS black holes \cite{Hawking:1982dh}).

Since the asymptotic observer can only measure the classical hair $M$, both the limits $\beta\to0$ and $\beta\to\infty$  in \cref{vacuumcontribmass} must lead to the same result, giving $c=1/\pi$.

For the asymptotic observer, the black-hole equilibrium temperature is $T_\text{H}$, so that \cref{vacuumcontribmass} gives 
\begin{equation}
N = \frac{\beta_\text{H} M}{2}=4 \pi G M^2.
\label{N}
\end{equation}

Seen by the distant observer, therefore, the number of oscillators scales holographically with the area of the event horizon. 
 The leading term of \cref{entropyexpanded}, together with \cref{N}, yields
\begin{equation}
S= N=4\pi G M^2
\end{equation}
which is exactly the BH  entropy. The sub-leading term in \cref{entropyexpanded} represents a logarithmic correction $N \ln T$, which is consistent with several results in the literature (see, e.g. Refs. \cite{Mann:1997hm, Akbar:2010nq, Carlip:2000nv, Mukherji:2002de, Medved:2004eh, Domagala:2004jt, Grumiller:2005vy, Fursaev:1994te, Setare:2006ww, Ghosh:1994wb, Gour:1999ta, Kaul:2000kf, Cadoni:2007vf, Gour:2003jj}).

\textit{Concluding Remarks}. All the information a distant observer can achieve about a SBH is its  mass and the response of the hole to perturbations, the QNMs spectrum. In this letter, we used this  fact to build a microscopic  description of the black hole in terms of a statistical ensemble of $N$ harmonic oscillators. In this way, we were able to derive, microscopically, the BH entropy as the leading term in the large-mass expansion. We found a subleading logarithmic correction to the BH entropy, in agreement with several results in the literature. In our microscopic description, the holographic character of the BH formula is  a natural consequence of the horizon-area scaling of $N$. An intriguing duality between black holes of small and large sizes also emerged in this approach.

\textit{Acknowledgments.} We thank Piero Olla and Edgardo Franzin for discussions.
\vfill

 \bibliography{SchwarzschildLetter27sett}
\bibliographystyle{ieeetr}

\end{document}